\documentclass[utf8]{frontiersFPHY} % for Physics and Applied Mathematics and Statistics articles

\setcitestyle{square} % for Physics and Applied Mathematics and Statistics articles
\usepackage{url,hyperref,lineno,microtype,subcaption}
\usepackage[onehalfspacing]{setspace}
\usepackage{graphicx}
\usepackage{amsmath}    % Advanced maths commands
\usepackage{amssymb}    % Extra maths symbols
\usepackage{txfonts}

\def\keyFont{\fontsize{8}{11}\helveticabold }
\def\firstAuthorLast{Lusso E.} %use et al only if is more than 1 author
\def\Authors{Lusso Elisabeta\,$^{1,*}$} %and others}
\def\Address{$^{1}$Dipartimento di Fisica e Astronomia, Universit\`a di Firenze, via G. Sansone 1, 50019 Sesto Fiorentino, Firenze, Italy \\ 
$^{2}$INAF -- Osservatorio Astrofisico di Arcetri, Largo Enrico Fermi 5, 50125 Firenze, Italy}
\def\corrAuthor{Corresponding Author}

\def\corrEmail{elisabeta.lusso@unifi.it}

\newcommand{\om}{\Omega_{\rm m}}
\newcommand{\ol}{\Omega_\Lambda}
\newcommand{\lo}{L_{2500}}
\newcommand{\lx}{L_{2\,\rm keV}}
\newcommand{\fo}{F_{2500}}
\newcommand{\fx}{F_{2\,\rm keV}}

\newcommand{\rev}[1]{{ #1}}

\begin{document}
\onecolumn
\firstpage{1}

\title[The eROSITA quasar Hubble diagram: a forecast]{Cosmology with quasars: predictions for eROSITA from a quasar Hubble diagram} 

\author[\firstAuthorLast ]{\Authors} %This field will be automatically populated
\address{} %This field will be automatically populated
\correspondance{} %This field will be automatically populated

\extraAuth{}% If there are more than 1 corresponding author, comment this line and uncomment the next one.
%\extraAuth{corresponding Author2 \\ Laboratory X2, Institute X2, Department X2, Organization X2, Street X2, City X2 , State XX2 (only USA, Canada and Australia), Zip Code2, X2 Country X2, email2@uni2.edu}

\maketitle

\begin{abstract}
%%% Leave the Abstract empty if your article does not require one, please see the Summary Table for full details.
The effort for understanding the matter and energy content of the Universe and its evolution relies on different probes, such as cosmic background radiation, cluster lensing, supernovae. Yet, we are still far from grasping what dark matter is made of, and what the physical origin of dark energy is. 
Our group has developed a technique that makes use of the observed non-linear relation between the ultraviolet and the X-ray luminosity in quasars to provide an independent measurement of their distances, thus turning quasars into {\it standardizable} candles. 
This technique, at present, it is mostly based upon quasar samples with data from public catalogues both in the X-rays and in the optical/ultraviolet and extends the Hubble diagram of supernovae to a redshift range still poorly explored ($z>2$). 
From the X-ray perspective, we are now on the eve of a major change, as the upcoming mission eROSITA is going to provide us with up to $\sim$3 millions of active galactic nuclei across the entire sky. 
Here we present predictions for constraining cosmological parameters such as the amount of dark matter ($\om$), dark energy ($\ol$) and the evolution of the equation of state of dark energy ($w$) through the Hubble diagram of quasars, based on the 4-year eROSITA all-sky survey. 
Our simulations show that the eROSITA quasars, complemented by redshift and broad-band photometric information, will supply the largest  quasar sample at $z<2$, but with very few objects available for cosmology at higher redshifts that survives the cut for the {\it Malmquist bias}, as eROSITA will sample the brighter end of the X-ray luminosity function. The power of the quasar Hubble diagram for precision cosmology lies in the high-redshift regime, where quasars can be observed up to redshift $\sim7.5$, essential to discriminate amongst different model extrapolations.
Therefore, to be competitive for cosmology, the eROSITA quasar Hubble diagram must be complemented with the already available quasar samples and dedicated (deep) large programmes at redshift $z>3$.
\tiny
 \keyFont{ \section{Keywords:} active galactic nuclei, quasar, observational cosmology, dark energy, surveys} %All article types: you may provide up to 8 keywords; at least 5 are mandatory.
\end{abstract}

\section{Introduction}
The driving forces behind the present era of precision cosmology have been the detection of anisotropies in the cosmic microwave background (CMB; e.g. \citealt{smoot1992}) and the discovery of the accelerated expansion of the Universe, based on the Hubble diagram (i.e. the distance modulus versus redshift relation) of Type Ia supernovae (SNe Ia), the standard candles {\it par excellence} \citep[e.g.][]{riess1998,perlmutter1999}. The currently accepted parameterization of our Universe is based on the so-called $\Lambda$ Cold Dark Matter ($\Lambda$CDM) model, hinging upon the existence of cold dark matter and on the cosmological constant ($\Lambda$). The nucleosynthesis of primordial elements, the large-scale galaxy distribution, and gravitational lensing are some of the usual probes into the nature of dark matter, and into how this interacts with visible (baryonic) matter. Yet, we are still far from grasping what the real constituents of this invisible cosmic ingredient are. Moreover, both the physical origin and the properties of dark energy are still unknown, as the interpretation of $\Lambda$ is plagued by the extreme degree of fine tuning required to obtain the right amount of dark energy observed today. Only the combination of multiple perspectives and of the optimal cosmological probes at different redshifts is the way forward to solve the dark matter and dark energy problems. 

In the past years, our group has developed a technique that makes use of the observed non-linear relation between the optical/ultraviolet and the X-ray luminosity in quasars \citep[e.g.][]{steffen06,just07,2010A&A...512A..34L,2016ApJ...819..154L}. In contrast to previous ideas of high stochasticity, this relation can be employed to standardize the emission of quasars \citep{rl15,lusso2019}. The methodology is complementary to the traditional resort to Type Ia SNe to estimate the cosmological parameters, yet it extends the Hubble diagram to a redshift range currently inaccessible to supernovae ($z=2-6$). 
\rev{The tightness of the UV-to-X-ray relation across both a wide redshift (up to $z\simeq5-6$) and luminosity range (see \citealt{2016ApJ...819..154L}) must be the manifestation of a universal physical mechanism that governs the disc-corona synergy in the quasar engines, yet the details of the physical process originating this relation is still unknown \citep[e.g.][]{2017A&A...602A..79L}.}
%A key consequence of this technique is that the optical/UV-to-X-ray luminosity relation must be the manifestation of a universal mechanism at work in the quasar engines, although the details on the physical process originating this relation are still unknown \citep[e.g.][]{2017A&A...602A..79L}. 

The main result of our work is that the distance modulus/redshift relation of quasars at $z < 1.4$ is in agreement with that of supernovae and with the concordance model \citep{rl15,lusso2019}. Yet, a deviation from the $\Lambda$CDM model emerges at higher redshift, with a statistical significance of 4$\sigma$. If we consider an evolution of the dark energy equation of state in form $w(z)=w_0+w_a\times z/(1+z)$, the data suggest that the dark energy density is increasing with time \citep{lusso2019,rl19}. 

In order to build a quasar sample that can be utilised for cosmological purposes, both X-ray and optical/UV data are required to cover the rest-frame 2 keV and the 2500\,\AA. At the time of writing, the most extended spectroscopic coverage in the optical/UV is provided by the \textit{Sloan Digital Sky Survey} \citep{2018A&A...613A..51P}, which supplies more than $\sim$500,000 quasars with spectroscopic redshift up to $z\sim7$. This sample needs to be cross-matched with the current X-ray catalogues, namely the {\it Chandra} CXC2.0\footnote{http://cxc.cfa.harvard.edu/csc2/} \citep{evans2010} and the 3XMM Data Release 8\footnote{http://xmmssc.irap.omp.eu/Catalogue/3XMM-DR8/3XMM\_DR8.html} \citep{rosen2016}, which contain \textit{all the X-ray sources detected by the XMM-Newton and Chandra observatories that are publicly available in the archives}. 
The number of quasars with a detection in both the UV and X-rays ranges from about a few thousands to $\sim$13,000, respectively. Once our filtering criteria are applied to select blue quasars with low levels of UV reddening and X-ray absorption and the Malmquist bias is corrected (the interested reader should refer to \citealt{rl15,2016ApJ...819..154L,bisogni2017FrASS,rl19} for details on the sample selection), the final samples drastically reduce to less than 2,000 objects ($\sim1000$ in the case of SDSS-CXC2.0, Bisogni et al. in preparation). 
Our leverage in building extended quasar samples for cosmology is thus entirely based upon archival data of pointed X-ray observations, which cover a very limited portion of the sky compared to SDSS, i.e. roughly 1000 deg$^2$ for both 3XMM-DR8 and CXC2.0 compared to $>$14,000 deg$^2$ for SDSS. 

We are now on the eve of the next major revolution in the field of X-ray astrophysics. The {\it extended Roentgen Survey with an Imaging Telescope Array} (eROSITA, \citealt{2012SPIE.8443E..1RP, 2012arXiv1209.3114M}) is the flagship instrument of the Russian {\it Spektrum-Roentgen-Gamma} (SRG) mission, and it will represent the most powerful and versatile X-ray observatory of the next decade. In the first 4 years of scientific operations, eROSITA will perform 8 deep scans of the entire sky, one every six months. When completed, the survey will be $\sim$20 times deeper than ROSAT at 0.5--2 keV, and it will provide the very first sensitive imaging of the whole sky in the hard band (2--10 keV). eROSITA will bring an improvement of over two orders of magnitude in the number of sources shining close to or above the break in the X-ray luminosity function (i.e. Fig. 5.2.2 in \citealt{2012arXiv1209.3114M}). eROSITA's sky will be dominated by the active galactic nuclei (AGN) population, with $\sim$3 million AGN expected by the end of the nominal 4-year all-sky survey at the sensitivity of $F_{0.5-2\,\rm keV} \simeq 10^{-14}$ erg s$^{-1}$ cm$^{-2}$ and with a median redshift of $z\sim1$. 

In this work we discuss the potential of the 4-year eROSITA all-sky survey for constraining cosmological parameters such as $\om$, $\ol$ and $w$, through the Hubble diagram of quasars.

\section{The simulated eROSITA quasar sample}

The SDSS-DR14 quasar catalogue contains 526,356 objects with $0.008<z<6.968$. We first selected a clean quasar sample in the optical/UV based on the selection criteria discussed in depth in our previous works \citep{2016ApJ...819..154L,rl15,rl19}. The main goal of this first step is to obtain the \textit{intrinsic} flux at the rest-frame 2500\,\AA. We excluded all quasars flagged as broad absorption line (BAL, BI\_CIV=0) and selected only the sources with a detection in all SDSS photometric bands, leading to a preliminary sample of 503,746 quasars. The SDSS-DR14 quasar catalogue also provides us with multi-wavelength information from several surveys, from the radio (FIRST survey) to the UV (GALEX survey; see section 7 in \citealt{2018A&A...613A..51P}). Thanks to this extended multi-band coverage, we built the spectral energy distributions (SEDs) that are then employed to compute the slope $\Gamma_{1}$ of a $\log(\nu)-\log(\nu F\nu)$ power law in the 0.3--1 $\mu$m (rest frame) range, and the analogous slope $\Gamma_{2}$ in the 0.3--0.145 $\mu$m range (rest frame). 
We assumed a standard SMC extinction law (\citealt{prevot84}, appropriate for unobscured quasars, \citealt{2004AJ....128.1112H,salvato09}) to estimate the $\Gamma_1-\Gamma_2$ correlation as a function of extinction (parametrized by E(B--V)). The $\Gamma_1-\Gamma_2$ value that corresponds to zero extinction (E(B--V)=0) is derived from the standard quasar SED of \citet[i.e. $\Gamma_1=0.82$, $\Gamma_2=0.40$]{2006AJ....131.2766R}. We then selected all objects within a circle centered at E(B--V)=0 with a radius of 0.8 (i.e. E(B-V)$<$0.1). From the SED we also calculated the flux at the rest-frame 2500\,\AA\ ($\fo$) and the one at 6 cm from the FIRST flux using a slope of $-0.8$, and we further excluded all the objects with $F_{6\,\rm cm}/\fo>10$. We also excluded all quasars in the sample defined as radio loud in the MIXR catalogue \citep{2016MNRAS.462.2631M} within a matching radius of 2 arcsec. This leads to a final clean sample in the UV of 291,944 quasars, within a redshift interval $0.061<z<5.25$ ($\langle z\rangle\simeq1.8$).

Since eROSITA is expected to survey the entire X-ray sky down to a flux that well matches the SDSS quasar optical magnitudes, we forecast that almost all SDSS quasars will be detected by eROSITA \citep{2016MNRAS.457..110M}. We thus simulated an X-ray flux measurement for each object as follows. 

\rev{We assumed the observed linear $\log \fx-\log \fo$ relation, with a slope $\alpha=0.6$, a flat $\Lambda$CDM cosmology with $H_0=70$ km s$^{-1}$ Mpc$^{-1}$, $\om=0.3$ and $\ol=1-\om$, and a dispersion in the $\fx-\fo$ relation on the order of $0.1$ dex. These values have been chosen to be representative of the mainstream models, although we know that the constraints on the cosmological parameters from observations in the local and in the early Universe are somewhat different (see Section~\ref{tension}). 
We started from fluxes as they are cosmology independent, and we have demonstrated in our previous works that the $\log \fx-\log \fo$ relation in narrow redshift bins displays the same slope (i.e. $\gamma\simeq0.6$) across a wide redshift range (see Figure~8 in \citealt{rl19} supplementary material). 
Moreover, the main aim of our simulations is to quantify the expected uncertainties on the cosmological parameters rather than focus on the absolute values per se. As such, we defer possible extensions of these simulations to non standard cosmological values and to a possible evolution of these parameters with redshift to future works.} 

As the eROSITA survey is flux limited, we also need to take into account the {\it Malmquist bias} (also known as the Eddington bias), which is a redshift dependent correction. 
We conservatively assumed an observed flux limit in the soft X-ray band of $3\times10^{-14}$ erg s$^{-1}$ cm$^{-2}$, which will be reached after the first year of operations, and considered all the sources with an expected monochromatic flux at 2\,keV that corresponds to about twice the value above, i.e. $5\times10^{-32}$ erg s$^{-1}$ cm$^{-2}$ Hz$^{-1}$, assuming a photon index of 1.9. This selection leaves very few sources at redshift higher than 2. We obtain a final sample of $\sim$11,000 quasars in the redshift range 0.061$-$2.850, with a mean redshift $\langle z\rangle\simeq0.64$, consistent with the predicted statistical properties based on the best available redshift-dependent AGN X-ray luminosity function \citep{2013A&A...558A..89K}.  

% -----------------
\begin{figure}[t!]
\begin{center}
\includegraphics[width=10cm]{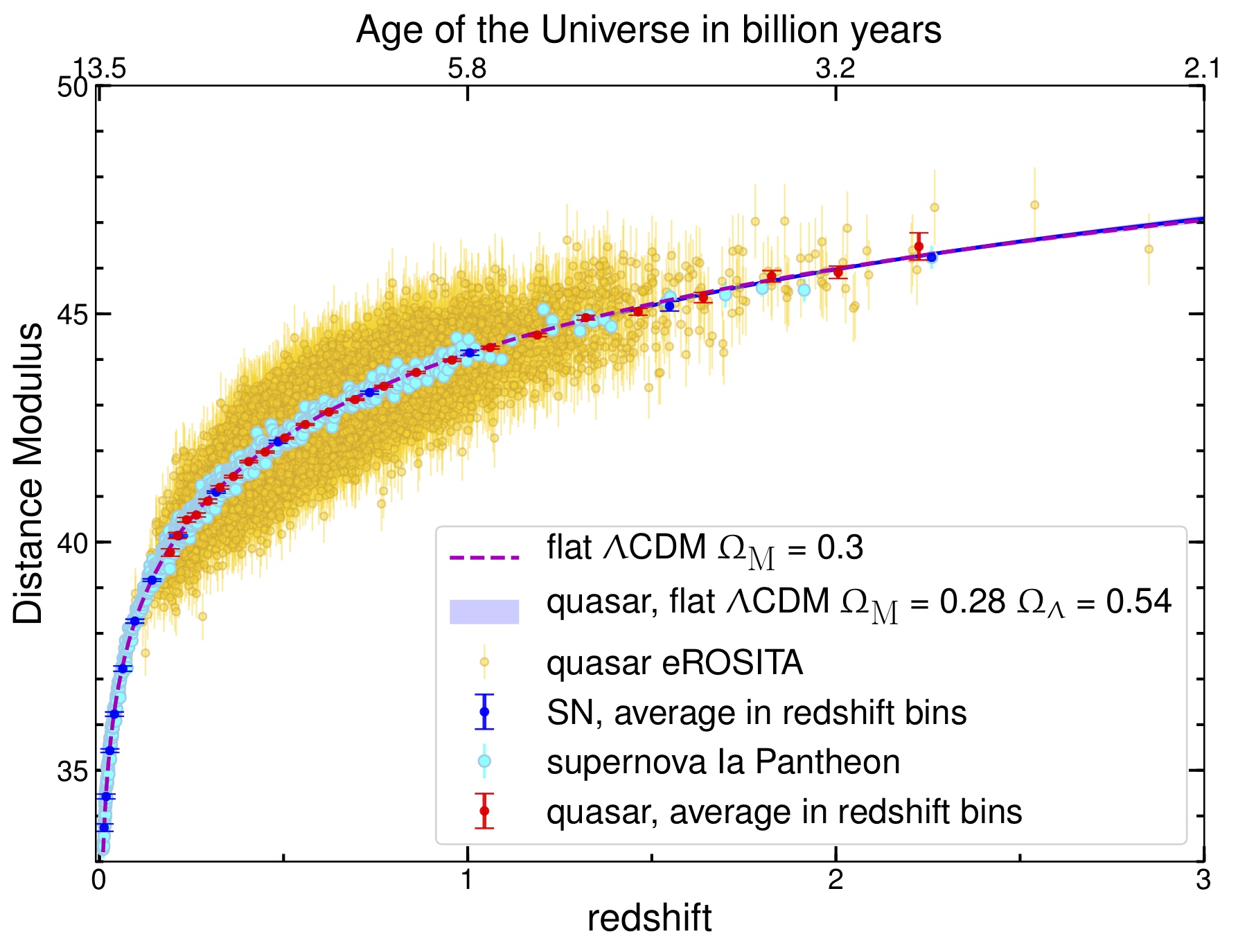}
\end{center}
\caption{Hubble diagram of eROSITA quasars ($\sim11,000$, yellow points) and SNe Ia ({\it Pantheon}, cyan points) with 1$\sigma$ uncertainties. Red points are the mean (also with 1$\sigma$ uncertainties) of the distance modulus in narrow redshift bins for quasars only (shown for visualisation purposes). The dashed magenta line shows a flat $\Lambda$CDM model fit with $\om=0.3$. The blue line is the best MCMC fit of the eROSITA quasars (with uncertainties) only for a $\Lambda$CDM cosmology where $\om$ and $\ol$ are left free to vary (\S~\ref{results}).
}\label{fig:hubble}
\end{figure}
% -----------------
\section{Results}
\label{results}

The distance modulus-redshift relation of the eROSITA quasars is presented in Figure~\ref{fig:hubble} with 1$\sigma$ uncertainties. \rev{The error bars on the distance modulus values for the quasar sample are estimated by propagating the uncertainties on the slope $\gamma$, $\fx$ and $\fo$. We assumed typical uncertainties on the slope and $\fx$ of 0.02 and $20\%$, respectively. Uncertainties on $\fo$ are computed by propagating the magnitude uncertainties from the SEDs we compiled for each SDSS quasar in the catalogue (see \S~2 in \citealt{rl19} supplementary material).} The red points are the mean (also with 1$\sigma$ uncertainties) of the distance modulus in narrow redshift bins for quasars (shown for visualisation purposes). Here we also show the SNe Ia sample from the {\it Pantheon} survey consisting of 1048 objects ranging from $0.01 < z < 2.26$ \citep{scolnic2018}. The dashed magenta line shows the input flat $\Lambda$CDM cosmology with $\om=0.3$ and $\ol=0.7$.

\begin{figure}[t!]
\begin{center}
\includegraphics[width=8.3cm]{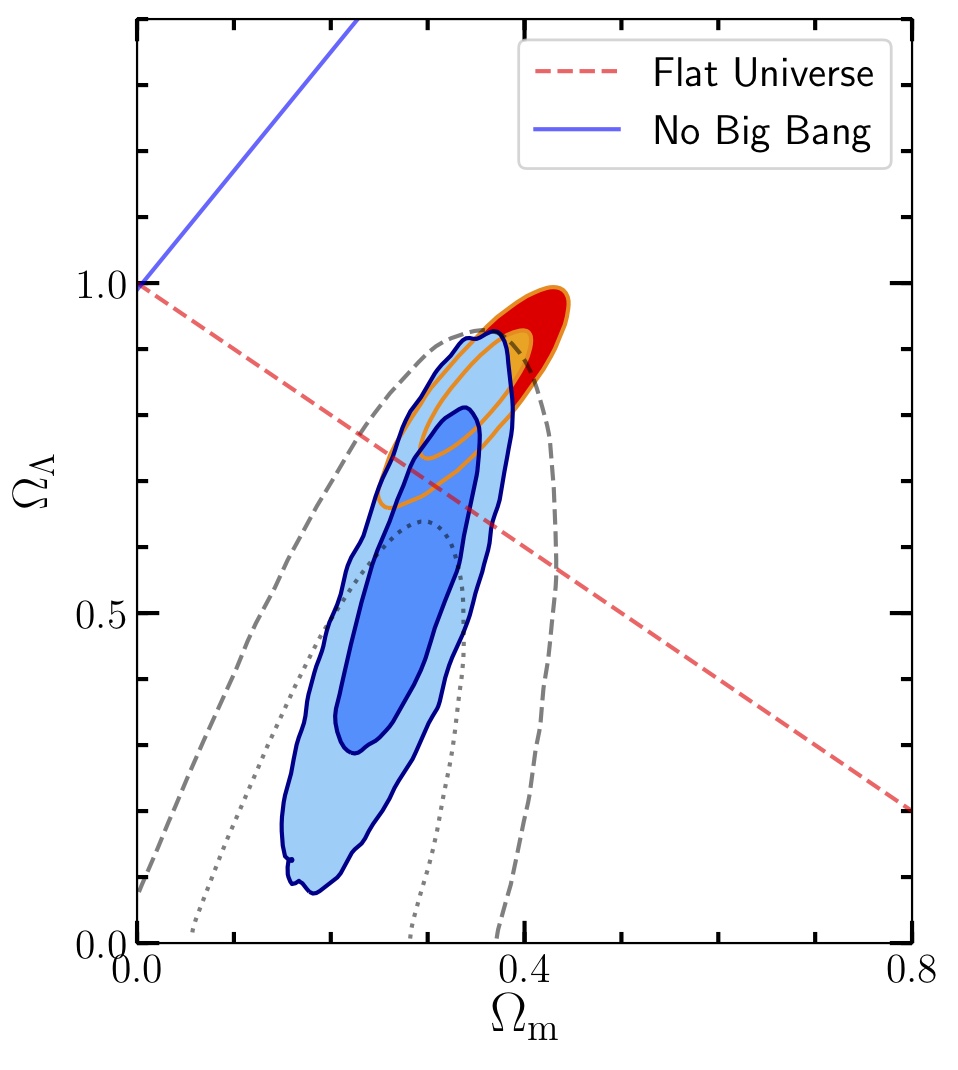}
\includegraphics[width=8.7cm]{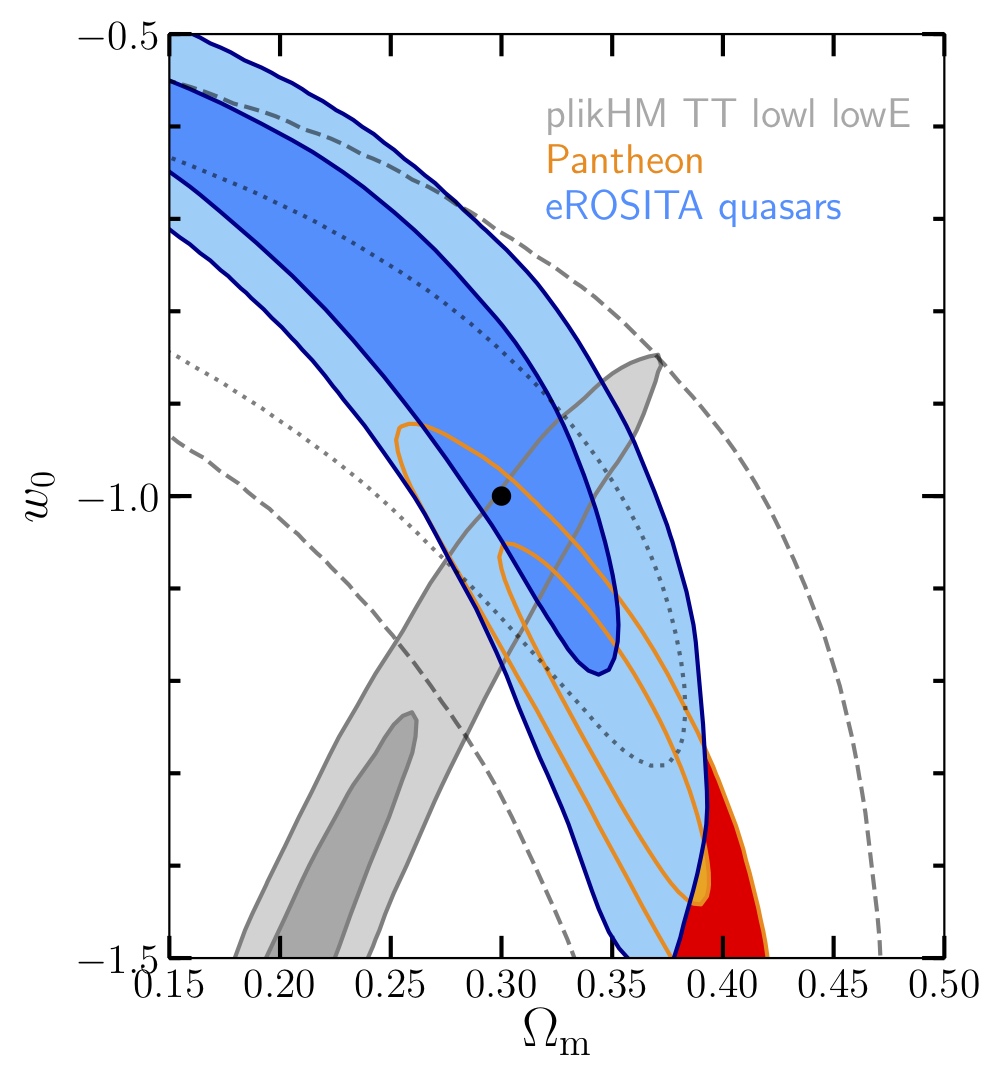}
\end{center}
\caption{{\it Left panel}: Confidence contours at 68\% and 95\% levels for the $\om$ and $\ol$ cosmological parameters in the $\Lambda$CDM model, where both $\om$ and $\ol$ are free to vary. All the plotted uncertainties are statistical. Orange/red contours: {\it Pantheon}. Blue/cyan: eROSITA simulated sample. {\it Right panel}: Confidence contours at 68\% and 95\% levels for the $\om$ and $w_0$ cosmological parameters in the $w$CDM model. All the plotted uncertainties are statistical. Orange/red contours: {\it Pantheon}. \rev{Filled} dark/light gray: {\it Planck} constraints from a Plik TT+low$l$+lowE likelihood. Blue/cyan: eROSITA simulated sample. The dashed/dotted contours represent the constraints on the cosmological parameters of a mock eROSITA quasar sample of $\sim17,000$ quasars assuming a dispersion in the $\lx-\lo$ relation of 0.2 dex.}\label{fig:res}
\end{figure}

We then fitted to this sample a $\Lambda$CDM model where $\om$ and $\ol$ are left free to vary, and a model with a dark energy equation of state parameter $w$ (assuming a flat Universe, $w$CDM). 
The results are shown in Figure~\ref{fig:res}, whilst a summary of the predictions on the cosmographic parameters from the analysis of the eROSITA quasar Hubble diagram is presented in Table~\ref{tab}. 
\rev{The confidence contours are at 68\% and 95\% levels and all the plotted uncertainties are statistical, computed from the marginalized posterior probability distributions.
All the simulations have been performed through a standard fully Bayesian procedure by making use of the affine invariant Monte Carlo Markov Chain ensemble sampler \citep{2013PASP..125..306F}. We adopted uniform priors on the cosmological parameters with $0<\om<1.2$ and $0<\ol<1.5$ for the $\Lambda$CDM model, whilst we used $0<\om<1.2$ and $-3<w_0<1$ for $w$CDM. 
The adopted likelihood also contains an intrinsic dispersion ($\delta$) as a free parameter (see \citealt{rl19} for details).
An additional free parameter is the cross-calibration ($\beta^\prime$) between supernovae Ia and quasars. Overall we have four free parameters for both the $\Lambda$CDM (i.e. $\om$, $\ol$, $\beta^\prime$ and $\delta$) and the flat $w$CDM (i.e. $\om$, $w_0$, $\beta^\prime$ and $\delta$).

Our method governs the shape of the Hubble diagram, whilst to determine the absolute value of the distances we need an external calibrator to build the distance ladder (like Cepheids are for supernovae Ia). 

As a consequence of the marginalization over the $\beta^\prime$, our technique does not provide any direct constraint on the Hubble parameter, $H_0$. In the simulations, the value of $H_0$ used for the calibration with the supernovae Ia is thus assumed and then marginalized over $\beta^\prime$. Considering a different value of $H_0$ would just modify the final cross-calibration with no change on the values of the best fit cosmological parameters and confidence intervals. 
}

% -----------------
% -----------------

% -------------
\begin{table}
\begin{center}
 \centering
  \setlength{\tabcolsep}{0.1 em}
\begin{tabular}{ccccc}
  \hline
\hline
  Model & $\om$ & $\ol$ & $w_0$ \\
  \hline
  $\Lambda$CDM  & $0.28\pm0.05$             & $0.54^{+0.17}_{-0.19}$ & ---  \\
  $w$CDM             & $0.26^{+0.07}_{-0.11}$ & --- & $-0.81^{+0.22}_{-0.28}$ \\
  \hline
\end{tabular}
\end{center}
\caption{Prediction on the cosmographic parameters from the analysis of the eROSITA quasar Hubble diagram.}
\label{tab}
\end{table}
% -------------

\section{Discussion}
The precision on $\om$ that can be achieved with the eROSITA quasars only is similar to that obtained today by supernovae Ia in the case $\om$ and $\ol$ are fitted simultaneously (i.e. $\om=0.35\pm0.04$, see Table 8 in \citealt{scolnic2018}). The current accuracy on $\ol$ from supernovae Ia is $\sim8\%$, whilst the precision on $\ol$ from the simulated quasar Hubble diagram is on the order of $\sim30\%$. This is due to the much higher dispersion of the data in the quasar Hubble diagram with respect to \textit{Pantheon} in the common redshift range. We also note that the assumed dispersion on the $\lx-\lo$ relation is rather optimistic. In fact, we can obtain a dispersion of $\sim0.12-0.15$ dex only when we consider pointed X-ray observations (see the Supplementary Material in \citet{rl19}), whilst the dispersion in the $\lx-\lo$ relation that we can achieve at present is $0.2-0.24$ dex. We also considered another mock quasar sample where we assume a dispersion of 0.2 dex, having similar statistics and redshift distribution to the one in Figure~\ref{fig:hubble}. The accuracy on $\om$ and $\ol$ decreases to $\sim$30\% and $\sim$40\%, respectively. 
The precision on $\om$ is more affected by the increased dispersion in the $\lx-\lo$ relation than the one on $\ol$. This is somewhat expected given the redshift distribution of the mock sample (see also Figure 5 in \citealt{2013A&A...558A..89K} for the statistical quasar sample predictions). The range of accuracy on $w_0$ for the best-case scenario and the more realistic one is shown in Figure~\ref{fig:res} for a $w$CDM. 

From our simulations one can conclude that, with the eROSITA quasars alone and the current observed dispersion in the $\lx-\lo$ relation, it will be challenging to provide stringent constraints on the cosmological parameters. Nonetheless, the simulated quasar sample does not include the hundreds of quasars at redshift $z>2.5$ that are already available from the public archives, which would not only improve the precision of the determination of both $\om$ and $\ol$, but will allow us to test with greater precision a possible evolution of the equation of state of dark energy with redshift $w(z)$. In fact, the parameter $\om$ is partly degenerate with $w_a$ in models with an evolving equation of state of the dark energy, $w_z$CDM. 

Even with a dispersion on the $\lx-\lo$ relation that matches the current ones (see Supplementary Fig. 6 in \citealt{rl19}), thanks to the much greater statistics at redshift lower than 2 offered by eROSITA, we will sample the {\it knee} of the Hubble diagram with several thousands of sources. The eROSITA sample combined to the high redshift quasars (in particular $z>3$) from the archives will allow us to test models where the dark energy equation of state $w$ is allowed to evolve with redshift. 

\subsection{On the tension of the Hubble parameter.}
\label{tension}
\rev{
It is clear from the left panel of Figure~\ref{fig:res} that CMB data alone do not constrain the equation of state of dark energy, $w$, due to strong geometrical model degeneracies. Indeed, {\it Planck} data on their own (i.e. CMB+lensing) can only assess the equation of state with $\sim$30\% uncertainty: $w=-1.57_{-0.40}^{+0.50}$, whilst this measure becomes $w=-1.04\pm0.1$ by considering a combination of {\it Planck} with Baryon Acoustic Oscillation (BAO, \citealt{planck2018}, {\it Planck} collaboration). This value is consistent with the expected one for a cosmological constant in the standard $\Lambda$CDM model, but the comparison between the concordance cosmological parameters obtained from the different probes brings out some tensions. For example, the most recent results from {\it Planck} assuming the standard $\Lambda$CDM cosmology are in tension at the 4.4$\sigma$ level with the direct measurement of the Hubble parameter ($H_0$, which measures the current expansion rate of the Universe) from Cepheids plus supernovae Ia \citep{riess2019}, and at about 2.5$\sigma$ with the matter density estimates from supernovae Ia (e.g. \citealt{rest2014}) and with the Ly$\alpha$ BAO measurements (e.g. \citealt{delubac2015,fontribera2014}). Recently, results on $H_0$ (assuming a standard flat $\Lambda$CDM cosmology) from 6 multiple-imaged quasar systems through strong gravitational lensing have confirmed this tension at more than 5$\sigma$ level with respect to {\it Planck} (H0LiCOW collaboration; \citealt{wong2019}). 
Whilst the reason for these tensions can be partially alleviated by accounting for the systematics in the different data sets (e.g. dependence of the supernovae Ia luminosity on age, \citealt{kang2020}), most of the discrepancy still remains unclear. In fact, within the $\Lambda$CDM framework, where $w$ is assumed to be constant ($w=-1$) across the cosmic time, there should be no difference between the $H_0$ value measured locally and the one measured in the early Universe. These discrepancies can in fact be the indication of new physics beyond the standard $\Lambda$CDM cosmology.

Our technique does not provide constraints on $H_0$ since this parameter is degenerate with the absolute cross-calibration of the Hubble diagram. Nonetheless, if we could confirm with high accuracy that $w$ indeed evolves with time (see \citealt{rl19,lusso2019}), this will provide an independent, compelling proof that this tension is real. Reducing the measurement uncertainties on cosmological parameters has become the main goal of current and forthcoming cosmological projects, in order to either corroborate the standard model or find new physics beyond it (see also results from N-body simulations, e.g. \citealt{adamek2016}).
Only the combination of different approaches, supported by an increased data quality and sample statistics, is the way forward to solve the $H_0$ tension. 

}

% precision achieved in the case we add high redshift? 

\section*{Funding}
We acknowledge financial contribution from the agreement ASI-INAF n.2017-14-H.O. 

\section*{Acknowledgments}
This work was initiated at the Aspen Center for Physics, which is supported by National Science Foundation grant PHY-1607611.

This research made use of Astropy a community-developed core Python package for Astronomy \citep{2018AJ....156..123A}. This research made use of Matplotlib, a 2D graphics package used for Python \citep{2007CSE.....9...90H}.

%\section*{Supplemental Data}
 %\href{http://home.frontiersin.org/about/author-guidelines#SupplementaryMaterial}{Supplementary Material} should be uploaded separately on submission, if there are Supplementary Figures, please include the caption in the same file as the figure. LaTeX Supplementary Material templates can be found in the Frontiers LaTeX folder.

%\section*{Data Availability Statement}
%The datasets [GENERATED/ANALYZED] for this study can be found in the [NAME OF REPOSITORY] [LINK].
% Please see the availability of data guidelines for more information, at https://www.frontiersin.org/about/author-guidelines#AvailabilityofData

%\bibliographystyle{frontiersinSCNS_ENG_HUMS} % for Science, Engineering and Humanities and Social Sciences articles, for Humanities and Social Sciences articles please include page numbers in the in-text citations
\bibliographystyle{frontiersinHLTHFPHY} % for Health, Physics and Mathematics articles
\bibliography{bibl}

%%% Make sure to upload the bib file along with the tex file and PDF
%%% Please see the test.bib file for some examples of references

%%% If you are submitting a figure with subfigures please combine these into one image file with part labels integrated.
%%% If you don't add the figures in the LaTeX files, please upload them when submitting the article.
%%% Frontiers will add the figures at the end of the provisional pdf automatically
%%% The use of LaTeX coding to draw Diagrams/Figures/Structures should be avoided. They should be external callouts including graphics.

\end{document}